\documentclass[sigplan]{acmart}

\copyrightyear{2026}
\acmYear{2026}
\setcopyright{cc}
\setcctype{by}
\acmConference[EuroMLSys '26]{Sixth European Workshop on Machine Learning and Systems}{April 27--30, 2026}{Edinburgh, Scotland Uk}
\acmBooktitle{Sixth European Workshop on Machine Learning and Systems (EuroMLSys '26), April 27--30, 2026, Edinburgh, Scotland Uk}
\acmDOI{10.1145/3805621.3807660}
\acmISBN{979-8-4007-2605-7/2026/04}

\AtBeginDocument{%
  }

\acmSubmissionID{75}

\usepackage{amsfonts}
\usepackage{algorithmic}
\usepackage{graphicx}
\usepackage{textcomp}
\usepackage{bmpsize}
\usepackage[table,xcdraw,dvipsnames]{xcolor}
\usepackage{enumitem}
\usepackage{lipsum}

\usepackage{xspace}
\usepackage{colortbl} 
\usepackage{multirow}
\usepackage{pifont}
\usepackage{booktabs}
\usepackage{adjustbox}
\usepackage{xstring} 
\usepackage{tikz}
\usepackage{xurl}  

\usepackage{float}
\usepackage[capitalize,nameinlink]{cleveref}
\crefname{section}{Sec.}{Secs.}
\Crefname{section}{Section}{Sections}

\crefname{subsection}{Sec.}{Secs.}
\Crefname{subsection}{Section}{Sections}

\crefname{figure}{Fig.}{Figs.}
\Crefname{figure}{Figure}{Figures}

\crefname{table}{Tab.}{Tabs.}
\Crefname{table}{Table}{Tables}

\crefname{equation}{Eq.}{Eqs.}
\Crefname{equation}{Equation}{Equations}

\usepackage{subcaption}
\usepackage{booktabs}
\usepackage{pifont}
\newcommand{\MakeMarkups}[3]{
  \expandafter\newcommand\csname #2\endcsname[1]{\textcolor{#3}{\textbf{[##1]}}}
}

\definecolor{darkgreen}{rgb}{0.4, 0.6, 0.2}
\MakeMarkups{Sina}{SA}{blue}
\MakeMarkups{Amir}{AM}{purple}
\MakeMarkups{Hamed}{HH}{magenta}
\MakeMarkups{Marios}{MK}{cyan}

\usepackage{listings}
\definecolor{clKeyword}{HTML}{0057B8}  
\definecolor{clComment}{HTML}{5C7E60}  
\definecolor{clString}{HTML}{B04300}   
\definecolor{clBG}{gray}{0.97}
\definecolor{clFrame}{gray}{0.65}

\lstdefinestyle{CCompact}{
  language=C,
  columns=fixed,            
  keepspaces=true,
  basicstyle=\ttfamily\scriptsize,     
  keywordstyle=\color{clKeyword}\bfseries,
  commentstyle=\itshape\color{clComment},
  stringstyle=\color{clString},
  numbers=left,
  numberstyle=\tiny\color{clFrame},
  stepnumber=1,
  xleftmargin=1.25em,       
  frame=single,
  framerule=0.3pt,
  rulecolor=\color{clFrame},
  backgroundcolor=\color{clBG},
  breaklines=true,
  breakindent=0pt,
  prebreak=\mbox{\textcolor{clFrame}{$\hookleftarrow$}},
  postbreak=\mbox{\textcolor{clFrame}{$\hookrightarrow$}},
  showstringspaces=false,
  aboveskip=4pt,
  belowskip=4pt,
  lineskip=1.2pt,
  tabsize=4
}

\usepackage{tikz}

\newcommand\eg{\emph{e.g.},\xspace}
\newcommand\ie{\emph{i.e.},\xspace}

\newcommand{\para}[1]{\noindent\textbf{#1.}}

\newcommand{\name}{\textit{AgenTEE}\xspace}
\newcommand{\names}{\textit{AgenTEE}'s\xspace}

\begin{document}

\title{\name: Confidential LLM Agent Execution on Edge Devices}

\begin{abstract}
Large Language Model (LLM) agents provide powerful automation capabilities, but they also create a substantially broader attack surface than traditional applications due to their tight integration with non-deterministic models and third-party services. While current deployments primarily rely on cloud-hosted services, emerging designs increasingly execute agents directly on edge devices to reduce latency and enhance user privacy. However, securely hosting such complex agent pipelines on edge devices remains challenging. These deployments must protect proprietary assets (\eg system prompts and model weights) and sensitive runtime state on heterogeneous platforms that are vulnerable to software attacks and potentially controlled by malicious users.

To address these challenges, we present \name, a system for deploying confidential agent pipelines on edge devices. 
\name places the agent runtime, inference engine, and third-party applications into independently attested confidential virtual machines (cVMs) and mediates their interaction through explicit, verifiable communication channels.
Built on Arm Confidential Compute Architecture (CCA), a recent extension to Arm platforms, \name enforces strong system-level isolation of sensitive assets and runtime state. 
Our evaluation shows that such multi-cVMs system is practical, achieving near-native performance with less than 5.15\% runtime overhead compared to commodity OS multi-process deployments.
\end{abstract}


\author{Sina Abdollahi}
\email{s.abdollahi22@imperial.ac.uk}
\orcid{0009-0008-3024-3106}
\affiliation{%
  \institution{Imperial College London}
  \city{London}
  \country{United Kingdom}
}
\author{Mohammad M Maheri}
\email{m.maheri23@imperial.ac.uk}
\affiliation{%
  \institution{Imperial College London}
  \city{London}
  \country{United Kingdom}
}

\author{Javad Forough}
\email{j.forough@imperial.ac.uk}
\affiliation{%
  \institution{Imperial College London}
  \city{London}
  \country{United Kingdom}
}
\author{Amir Al Sadi}
\email{a.al-sadi@imperial.ac.uk}
\affiliation{%
  \institution{Imperial College London}
  \city{London}
  \country{United Kingdom}
}

\author{Josh Millar}
\email{joshua.millar22@imperial.ac.uk}
\affiliation{%
  \institution{Imperial College London}
  \city{London}
  \country{United Kingdom}
}

\author{David Kotz}
\authornote{This work was performed while Professor Kotz was in residence at Imperial College London.}
\email{David.F.Kotz@dartmouth.edu}
\affiliation{%
  \institution{Dartmouth College}
  \city{Hanover}
  \country{NH, USA}
}
\author{Marios Kogias}
\email{m.kogias@imperial.ac.uk}
\affiliation{%
  \institution{Imperial College London}
  \city{London}
  \country{United Kingdom}
}
\author{Hamed Haddadi}
\email{h.haddadi@imperial.ac.uk}
\affiliation{%
  \institution{Imperial College London}
  \city{London}
  \country{United Kingdom}
}



\renewcommand{\shortauthors}{Abdollahi et al.}




\maketitle

\section{Introduction}

Recent advances in Large Language Models (LLMs) have enabled a new class of software, LLM agents\footnote{In this paper we use \textit{agents} and \textit{LLM agents} interchangeably.}, that can autonomously reason over user instructions, plan multi-step tasks, and interact with external services.
An agent combines a language model with a runtime that maintains context, interprets user requests, selects relevant services, invokes them, and integrates their outputs into subsequent reasoning steps.

Traditionally, agents are deployed as cloud-hosted services, often granted access to users’ data through APIs or synchronization mechanisms~\cite{Usingtools,Microsoft365Copilothub}.
In contrast, emerging edge-device designs execute both the agent runtime and the underlying model locally. Such deployments improve privacy and data locality by confining users' personal data to the device, giving users greater control over its exposure to external services~\cite{wu2024isolategpt,debenedetti2025defeating,chen2025octo}. 

Compared to traditional software, agents face a significantly broader attack surface. On one hand, they require extensive access to third-party services and user's data to perform complex tasks effectively. These accesses expose the agent to untrusted or potentially malicious inputs (\eg emails or documents) that enter its processing pipeline. On the other hand, their core reasoning engine---the LLM itself---cannot reliably distinguish trusted system instructions from untrusted inputs, making it inherently vulnerable to attacks such as data exfiltration, unintended action execution, or safeguard bypass. As a result, LLM agents demand stronger isolation guarantees than conventional applications.



Current agent deployments on edge devices primarily rely on OS mechanisms---such as multi-processing and system-call filtering (\eg seccomp~\cite{seccomp})---to isolate agent components from the rest of the software stack and from co-resident applications~\cite{wu2024isolategpt,debenedetti2025defeating}. 
This is typically adequate for low-stakes, simple agents (\eg an agent retrieving public news from the internet). 
However, once an agent’s workflow includes \textit{proprietary assets} (\eg specialized models or confidential agent code), software-only isolation provides inadequate protection for sensitive runtime state and data-in-use, since these assets must remain confidential even from privileged platform software.

The problem becomes even more challenging where the agentic service  components needs to be provided by \textit{mutually distrustful stakeholders}. 
For example, one party may supply the runtime while another supplies the model, alongside third-party integrated applications. 
%

What is fundamentally required is an execution environment that can simultaneously host components of agentic workloads (\ie agent, model, and third-party applications) and protect them from both the hosting platform and from each other.
The execution substrate must prevent the commodity OS from inspecting sensitive runtime state and messages exchanged between these components, while enabling interaction without exposing proprietary assets.
%

Trusted Execution Environments (TEEs) provide a hardware-backed foundation for such a execution environment, offering protected execution and data-in-use confidentiality even against privileged software. 
In contrast to cryptographic-based approaches (\eg~\cite{riazi2018chameleon,gilad2016cryptonets,van2019sealion,maheri2025telesparse,maheri2025zk}), TEEs are often more practical for end-to-end systems, as they preserve near-native performance relative to conventional execution environments.

Despite these advantages, widely deployed TEE architectures such as Arm TrustZone~\cite{trustzone} are a poor fit for isolating multi-component agent pipelines. They are primarily designed for small, vendor-specific services, such as authentication (\eg Face ID~\cite{appleencalve}) and digital rights management (\eg Widevine~\cite{Widevine}), and face significant limitations when supporting complex, general-purpose workloads~\cite{siby2024guarantee}.

Recent extensions to Arm architecture help address this gap. 
Arm Confidential Compute Architecture (CCA)~\cite{ccasite} enables general-purpose confidential virtual machines (cVMs), called \textit{realms}, that execute in hardware-isolated memory, protected from the hosting operating system and hypervisor. Realms can run unmodified software and can do not face any inherit memory size limitation.
Each realm is cryptographically measured at launch and supports remote attestation, allowing mutually distrustful components to verify identity and integrity before provisioning secrets or exchanging sensitive data.

In this paper, we argue that secure agentic systems executing on edge devices and handling proprietary assets fundamentally require hardware-backed confidential execution. 
We present \name, a framework that deploys the agent runtime, LLM inference engine, and third-party applications into independently attested realms, and mediates their interaction through explicit, verifiable communication channels. 
By confining proprietary assets and intermediate runtime state to hardware-protected memory, \name enables secure composition of mutually distrustful components on a single device with strong isolation and minimal performance overhead.



%
%

This paper makes the following \textbf{contributions}:
\begin{itemize}
    \item We identify the security requirements of edge-device LLM agents and explain how Arm CCA satisfies these requirements.
    \item We design \name, a system for executing agent runtime and third-party applications within confidential execution environments, enabling mutual attestation and hardware-enforced isolation between components.
    \item We implement\footnote{Our evaluation artifacts is available at \url{https://agentee-paper.github.io/}} and evaluate \name on Arm CCA, demonstrating that confidential edge-device LLM agents are practical and scalable on modern edge hardware, incurring less than 5.15\% overhead compared to native execution.
\end{itemize}

\section{Background \& Motivation}
In this section, we first provide background on LLM agents (\cref{sec:llmagent}) and discuss their system-level isolation requirements (\cref{sec:agentsecurity}). We then review existing TEEs for edge devices and position Arm CCA as a promising foundation for \name (\cref{sec:TEEonedge}).
\begin{figure}[t]
    \includegraphics[width=\linewidth]{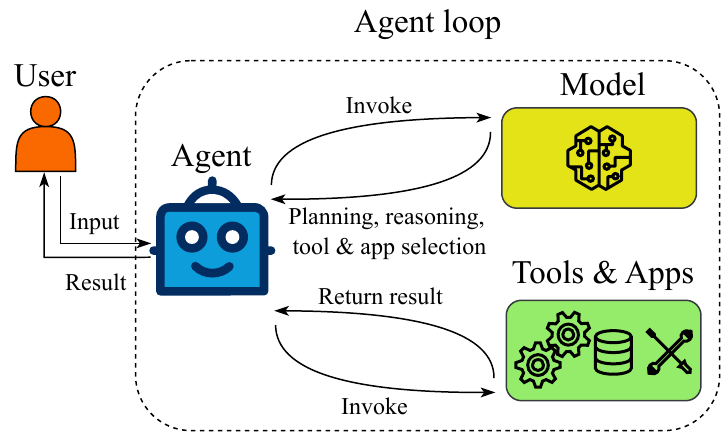}
    \caption{LLM agent architecture}
    \label{fig:LLMagent}
    \vspace{-15pt}
\end{figure}
\subsection{LLM Agents} \label{sec:llmagent}

\Cref{fig:LLMagent} illustrates an LLM agent flow. An LLM agent integrates a language model with a runtime that maintains context, interprets user requests, selects relevant components, invokes them, and integrates their outputs into subsequent reasoning steps. 
%
%
The LLM serves as the agent’s core reasoning component, interpreting natural language instructions, generating plans, and selecting appropriate tools to invoke.
Tools and third-party applications broaden the functional scope of the agent by enabling access to real-time information~\cite{yao2022react,zhao2024attacks}, conducting complex calculations~\cite{schick2023toolformer}, and executing specialized tasks such as image recognition~\cite{qin2023toolllm}.

\para{Agent-Model Interaction}
Modern LLMs are trained to interpret role-labeled inputs, primarily the \textit{system prompt} and the \textit{user prompt}, and to prioritize the system prompt during reasoning and output generation~\cite{anthropic_context_engineering}.
A central responsibility of an agent is to construct and properly label the model input.
The agent uses the user prompt to convey user requests, interaction history, and external input data, which may be untrusted or adversarial, while the system prompt encodes trusted instructions that govern the model’s behavior, including its role, tone, safety constraints, and operational policies~\cite{vertex_system_instructions,anthropic_context_engineering}.
The system prompt further defines accessible tools and external resources, as well as the conditions under which they may be invoked.




\subsection{Isolation Requirements of Agent Workflow}\label{sec:agentsecurity}
In this section, we discuss several internal assets whose confidentiality and integrity are fundamental to the secure operation of an agent. 
These assets collectively form the agent’s control plane and runtime state. 
We argue how leakage or unauthorized modification at any of these layers can undermine the agent’s runtime security and reliability.

\para{Agent}
Agent code---including prompt templates (\eg system prompt), decision logic, orchestration rules, and memory management---is central to an agentic system’s reliability and policy compliance, and may treated as proprietary~\cite{hui2024pleak}. 
Organizations invest significant effort in engineering these components to achieve efficiency, enforce policies, guide tool usage, and achieve safe behavior, making them a source of competitive advantage that requires protection from unauthorized access. 

Prior works show that even partial leakage of these components significantly increases the feasibility of indirect and tailored prompt injection attacks~\cite{greshake2023not,agarwal2024prompt}, data exfiltration~\cite{qi2024follow,clouddataexfilteration}, and secret leakage~\cite{devinsecret}. This observation highlights that protecting internal agent logic is not merely an intellectual property concern, but a fundamental security requirement.

\para{Inference Engine}
The inference engine is the runtime component responsible for executing a trained model to produce outputs from input prompts. 
In doing so, it handles two confidentiality-critical assets: (1) the model weights, which encode the model’s learned capabilities and behavior, and (2) transient runtime state, most notably the key--value (KV) cache, which stores intermediate attention states to accelerate LLM inference~\cite{kwon2023efficient}. 
Protecting both assets is essential, since their exposure or manipulation can directly affect security and correctness.

Model weights carry substantial economic and operational value, reflecting significant investment in training and fine-tuning; proprietary weights therefore require strong confidentiality guarantees. 
Even when weights are public, integrity during inference remains crucial: LLMs can be highly sensitive to parameter-level perturbations, where a small fraction of parameter/feature outliers can disproportionately influence model quality~\cite{dettmers2022gpt3,an2025systematic}, and pruning even a single critical parameter can catastrophically degrade quality~\cite{yu2024super}.
\begin{figure}[t]
    \includegraphics[width=0.9\linewidth]{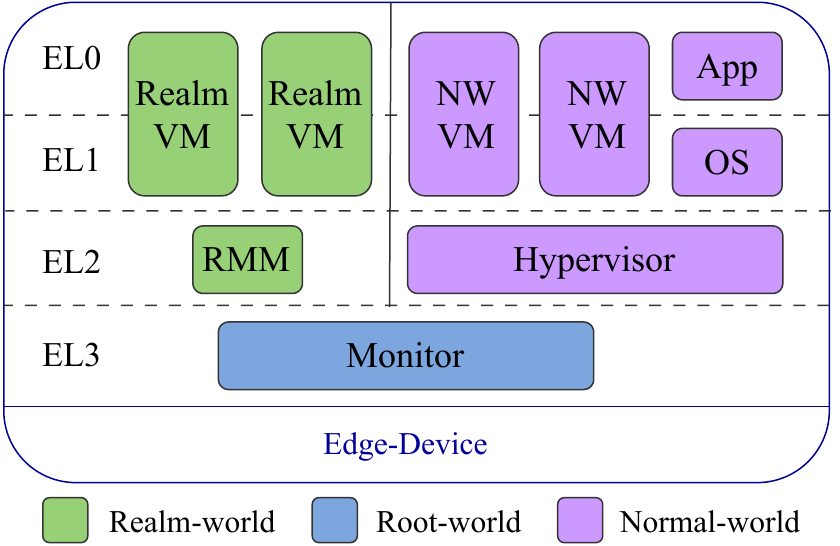}
    \caption{Arm Confidential Compute Architecture}
    \label{fig:ArmCCA}
\end{figure}
Beyond weights, the inference engine’s runtime state also constitutes a security-sensitive asset. 
The primary example is the key--value (KV) cache~\cite{luo2025shadow,wu2025know}, which -- while introduced for performance -- encodes semantic representations of previously processed context. 
If an adversary can read or tamper with cached key--value states, system prompts or privileged instructions may be reconstructed or implicitly reactivated; moreover, cache manipulation can steer behavior or trigger policy bypass without changing the visible prompt. 
Accordingly, the KV cache requires both confidentiality and integrity protection.
\para{Third-Party Applications}
Third-party applications frequently require access to sensitive credentials, such as API keys, database secrets, or authentication tokens. They may also employ business logic to their local services. These secrets and business logic must be consistently protected against adversaries, including malicious applications running on the same device.

\subsection{TEEs on the Edge}\label{sec:TEEonedge}
%

Historically, edge-device TEEs such as Arm TrustZone~\cite{trustzone} and Intel TDX~\cite{intelsgx} have been used to protect sensitive applications, including key management, digital rights management (DRM), federated learning~\cite{mo2021ppfl}, and machine learning inference~\cite{sun2023shadownet}. 
However, edge-device TEEs have their own limitations, including restricted  memory size \cite{siby2024guarantee} and limited accessibility for third-party developers \cite{abdollahi2025early,kuhne2024aster}. Such constraints complicate the deployment of complex, memory-intensive workloads, such as LLM agents.

To mitigate these limitations, Android introduced the Android Virtualization Framework (AVF)~\cite{avfarchitecture}, which enables isolated VMs protected from the Android kernel. However, these VMs are not protected by hardware-enforced mechanisms and their security relies on the Android hypervisor, which remains part of the trusted computing base (TCB) of the VM\@.

\para{Arm CCA} Arm CCA is an extension to the Armv9-A architecture targeting both edge devices and cloud servers.
As illustrated in \Cref{fig:ArmCCA}, CCA introduces a new  execution state called the \textit{realm-world}, alongside the traditional normal-world (NW). 
CCA enables cVMs, referred to as realm VMs (or simply realms), which execute within the realm-world. 
Realm-world memory is protected through hardware-enforced isolation mechanisms, making it inaccessible to entities operating in the normal-world (including the OS and hypervisor).
Consequently, even a compromised host operating system or hypervisor cannot directly access or modify realm memory or virtual CPU state.

In CCA, the hypervisor retains responsibility for resource management, including CPU scheduling and memory allocation. However, hypervisor operations on realm resources are mediated by a trusted lightweight firmware component called the Realm Management Monitor (RMM). By design, CCA significantly reduces the TCB: the hypervisor (approximately 300K lines of code) is excluded from the TCB, and security enforcement is delegated to the comparatively small RMM (approximately 27K lines of code).

\para{CCA provides appropriate abstraction}
CCA introduces several properties that make it a suitable foundation for \name. Firstly, CCA enables flexible, general-purpose confidential computation by moving beyond fixed-function TEEs (\eg TrustZone) to support general purpose realms, within a single device, making it a natural fit for \name. 
Unlike TrustZone, which relies on statically partitioned memory, CCA allows realms to be instantiated dynamically as long as sufficient RAM is available. These realms can execute existing software stacks with no modification, even if they were originally designed for a conventional execution environment.
This supports more complex, multi-component agent pipelines---spanning the runtime, model, and third-party services---and makes it practical to deploy components from heterogeneous trust domains on commodity edge devices without specialized hardware configurations.
%

%

Second, CCA’s RISC-based architectural design enables extensibility and customization. 
Its minimal and modular design allows incremental extensions to support emerging workload requirements without redesigning the TCB. 
Prior work has demonstrated extensions for efficient inter-realm communication and protected data sharing~\cite{abdollahi2025confidential}, as well as confidential GPU acceleration~\cite{sang2025portal,sridhara2024acai}, enabling practical deployment of multi-realm and performance-sensitive workloads.

%

\section{\name Overview}
In this section, we provide an overview of \name, its threat model and pipeline.
\begin{figure}[t]
    \includegraphics[width=\linewidth]{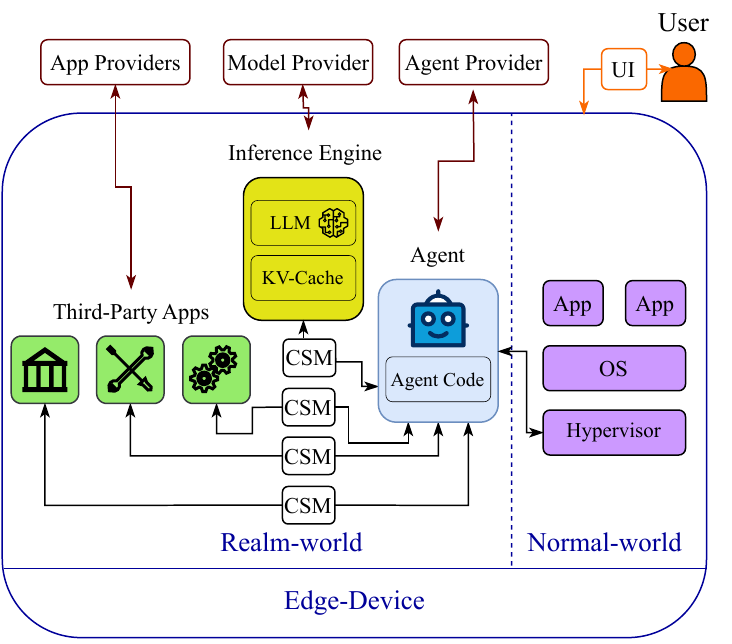}
    \caption{\names system architecture}
    \label{fig:agenTEE}
\end{figure}

\subsection{System and Security Model}

\para{System Model} 
\name involves five major entities: \textit{model providers}, \textit{agent providers}, \textit{application providers}, the \textit{users}, and the \textit{users’ device}. 
Model providers train and distribute machine-learning models, agent \allowbreak providers implement the agent code, which are both proprietary.
Application providers offer external services invoked by the agent (\eg banking, travel booking, or other cloud services).
The users' devices are Armv9-A platform supporting the CCA architecture, where the normal-world runs a commodity operating system (\eg Android) that controls hardware resources, user interfaces (UI), and conventional applications, while also supporting the execution of cVMs in the realm-world.

\para{Trust Model} 
We assume that users trust their own devices and the normal-world environment through which inputs are received and delivered to the agent\footnote{More complex scenarios in which the NW lies outside the user’s trust domain are out of scope for this work.}. In contrast, the agent provider, model provider, and third-party application providers do not trust one another, nor do they trust the NW running on users’ devices.
All stakeholders trust the underlying CCA hardware and firmware (\ie the RMM and Monitor in \cref{fig:ArmCCA}) and assume that its architectural security guarantees hold.

\para{Threat Model} 
We assume that  model provider, agent \allowbreak provider, and third-party application provider, and NW stack may attempt to obtain information about other parties’ proprietary assets. Such assets include model weights, agent runtime code or APIs provided by third-party applications. 
Physical attacks and microarchitectural side-channel attacks against the TEE hardware are out of scope. We also assume no compromise of cryptographic primitives or violations of CCA’s architectural guarantees. 

\subsection{\name Pipeline}
\Cref{fig:agenTEE} shows the system architecture of \name. \name organizes the edge-device agent pipeline entirely within the realm world. The agent runtime, inference worker, and third-party applications are each deployed in separate cVMs. All interactions between these components occur through protected realm memory, preventing the host operating system or hypervisor from observing or manipulating sensitive runtime state. 

 The pipeline begins with each stakeholder deploying its component into a realm using the standard CCA initialization flow, which loads a publicly available image into realm memory. Upon launch, each realm establishes a secure network connection (TLS) with the respective owner, providing an RMM-signed attestation token. This token serves as cryptographic evidence that the realm is executing the software stack each owner expects. Once each owner verifies its realm, it uses the secure network connection to transmit proprietary assets (\eg model, prompt, or agent runtime code) to its realm. 

To enable efficient and mutually authenticated inter-cVM communication, \name integrates CAEC~\cite{abdollahi2025confidential} design, which provides Confidential Shared Memory (CSM) between realms. CSM enables hypervisor's inaccessible channels, allowing peer realms to exchange data through memory regions inaccessible to the normal world. \name adopts CAEC’s inter-realm attestation protocol to ensure that communication occurs only between verified and authorized realms.

Once the agent realm establishes a trusted channel with the model realm, it signals readiness to the normal world. A user-interface application in the normal-world mediates user interactions by forwarding requests to the agent realm, while remaining unable to access or inspect the protected internal state of the agent or the model.

\para{Security Analysis}
\name mitigates the threats outlined in our model  by deploying the agent runtime, inference engine, and third-party applications in separate realms. \name confines proprietary assets ( and sensitive runtime state  to hardware-isolated memory that is inaccessible to the normal world, including a compromised OS or hypervisor. 
Remote attestation ensures that each stakeholder provisions its assets only to a verified realm running the expected software stack, preventing unauthorized modification or impersonation.

\vspace{6pt}

\section{Implementation}
There is no native Arm CCA hardware yet commercially available. We therefore rely on OpenCCA~\cite{bertschi2025opencca}, an open-source prototype implementation of Arm CCA. OpenCCA implements a prototype of CCA on an embedded cost-effective hardware, Radxa Rock 5B~\cite{ROCK5B}.
Although this prototype cannot exactly forecast system performance on future CCA hardware, it provides a practical and best-effort approximation in the absence of commercial hardware support.

\para{Software Stack}
Our implementation builds on Trusted Firmware-A~\cite{TF-A} (v2.11) as the Monitor and the reference RMM implementation~\cite{RMM} (v0.5.0). We use linux-cca~\cite{linux-cca} (v5+v7) as both host and guest operating system, and kvmtool-cca~\cite{kvmtool-cca} (v3/cca) as the virtual machine manager (VMM). To enable CSM between realms, we apply CAEC’s patches to the RMM, linux-cca, and kvmtool-cca used in the implementation.
\begin{table*}[!t]
\centering
\caption{Inference and end-to-end latency (second) for chatbot and itinerary agents under different mechanisms.}
\label{tab:combined_agents}
\footnotesize
\setlength{\tabcolsep}{6pt}

\begin{tabular}{lllcccc}
\toprule
\multirow{2}{*}{\textbf{Agent}}
& 
& 
& \multicolumn{2}{c}{\textbf{GPT2-Medium-q8\_0 (437MB)}}
& \multicolumn{2}{c}{\textbf{Llama-3.2-1B-Instruct-Q4\_0 (773MB)}} \\
\cmidrule(lr){4-5} \cmidrule(lr){6-7}
& & & \textbf{Inference} & \textbf{End-to-End}
& \textbf{Inference} & \textbf{End-to-End} \\
\midrule

\multirow{5}{*}{Chatbot}
& \multirow{3}{*}{Isolation} & NW Process         & 92.16  & 93.45  & 277.79 & 277.80 \\
&                            & NW VM              & 96.26  & 96.29  & 284.89 & 284.90 \\
&                            & \name              & 98.22  & 98.25  & 289.52 & 289.55 \\
\cmidrule(lr){2-7}
& \multirow{2}{*}{Overhead}  & \name\ vs NW Process & 6.57\% & 5.14\% & 4.22\% & 4.23\% \\
&                            & \name\ vs NW VM      & 2.04\% & 2.04\% & 1.63\% & 1.63\% \\
\midrule

\multirow{5}{*}{Itinerary}
& \multirow{3}{*}{Isolation} & NW Process         & 163.50 & 163.82 & 462.30 & 462.31 \\
&                            & NW VM              & 168.98 & 168.99 & 467.92 & 473.26 \\
&                            & \name              & 170.76 & 170.77 & 481.25 & 485.18 \\
\cmidrule(lr){2-7}
& \multirow{2}{*}{Overhead}  & \name\ vs NW Process & 4.44\% & 4.24\% & 4.10\% & 4.08\% \\
&                            & \name\ vs NW VM      & 1.05\% & 1.05\% & 2.85\% & 2.52\% \\
\midrule
\end{tabular}
\end{table*}

\para{User-Space CSM Interface}
To facilitate application-level communication over CSM, we design a lightweight (184 line of code) Python module that abstracts CSM usage for user-space components. While CAEC~\cite{abdollahi2025confidential} provides a Linux driver that exposes CSM regions to user space, it does not define how applications should structure communication over this shared memory. Our module builds on the exposed CSM regions and partitions each inter-realm CSM region into multiple logical half-duplex channels, enabling structured and efficient message passing between components executing in separate realms.

\para{Implemented Agents} To evaluate \name under realistic edge-device agent workloads, we implement two representative LLM agents that share the same execution architecture but differ in prompt structure and computational intensity. We implemented both agents using LangChain framework~\cite{LangChain}.

\textit{Chatbot Agent:}
The chatbot agent represents a lightweight conversational assistant. It accepts free-form natural language queries, constructs a short prompt using a fixed system prompt, and forwards it to the LLM for response generation. This agent models common edge-device assistants used for question answering, summarization, and explanation, and serves as a baseline workload with modest communication and inference demands.

\textit{Itinerary Planning Agent:}
The itinerary agent represents a task-oriented planning workload. Given a destination, trip duration, budget, interests, and additional constraints, the agent constructs a structured prompt that requests multi-day schedules, activity lists, and cost estimates. Compared to the chatbot agent, this workload produces longer prompts and responses and exercises both the communication channel and LLM inference more heavily.

 \vspace{3pt}

\section{Preliminary Evaluation}\label{sec:endtoendoverhead}
To evaluate the performance overhead of \name, we measure the dominant cost of an agent workflow: the agent–model interaction. Specifically, we measure the end-to-end latency of each agent–model query under different isolation scenarios to quantify the cost of hardware-backed isolation in \name.

We conduct experiments using two representative agents (\ie a chatbot and an itinerary planner), two models (GPT2-Medium~\cite{gpt2-medium} and Llama-3.2-1B~\cite{llama1b}), and three isolation configurations.
The isolation scenarios are as follows:(1) \name, where two cVMs communicate through CSM, assuming the entire NW is untrusted; (2) two NW VMs communicating via shared memory, where the NW hypervisor is trusted; and (3) two NW processes communicating through shared memory within the NW OS (baseline), where the NW OS and hypervisor are trusted. These configurations progressively relax the trust assumptions and represent decreasing levels of isolation.
For each configuration, we measure both the inference latency within the inference engine and the end-to-end latency observed by the agent when issuing a model query.

\Cref{tab:combined_agents} reports the median results over nine typical queries. The end-to-end overhead of \name compared to NW VMs remains below 2.53\% across all settings, while the end-to-end overhead relative to NW processes is below 5.15\%. These results demonstrate that \name introduces minimal performance overhead despite providing stronger isolation guarantees, remaining close even to the weakest isolation baseline, which is inter-process communication within the NW OS.

We note that our experiments are conducted on a resource-constrained prototype platform that is less powerful than modern edge devices (\eg smartphones). On contemporary mobile CPUs, we expect the absolute latency to be much less. In addition, upcoming CCA extension~\cite{armCCAupdate} will support secure assignment of specialized accelerators (\eg GPUs or NPUs) to realms, enabling the \names inference engine to leverage hardware acceleration for token generation, improving absolute latency.


 \section{Related Work}
 \para{LLM Agents}
Recent work explores the deployment of LLM agents on edge devices~\cite{wu2024isolategpt,debenedetti2025defeating,wang2024mobile}. Prior studies recognize internal agent state—such as system prompts—as confidential assets whose exposure enable attacks including indirect prompt injection, data exfiltration, and tool manipulation~\cite{greshake2023not,agarwal2024prompt,devinsecret,qi2024follow,clouddataexfilteration}.
Beyond prompt manipulation, model-centric threats target the integrity and confidentiality of model weights and runtime state: LLMs are highly sensitive to parameter perturbations~\cite{dettmers2022gpt3,an2025systematic,yu2024super}. 

\para{Confidential Computing}
Confidential computing research investigates hardware-backed isolation mechanisms to protect machine learning workloads, including TEEs such as Arm TrustZone~\cite{trustzone,brasser2019sanctuary,sun2022leap}. More recently, Arm CCA has gained attention for extending TEE capabilities to support general-purpose confidential virtual machines, with several works exploring performance, accelerator integration, and system-level enhancements~\cite{shen2022soter,sridhara2024acai,wang2024cage,sang2025portal}. CAEC~\cite{abdollahi2025confidential} further introduces CSM between realms, enabling efficient data sharing across isolated cVMs and improving the practicality of multi-cVM deployments.

\section{Conclusion}
In this paper, we presented \name, a system for confidential edge device deployment of LLM agent pipelines. By leveraging Arm CCA, \name isolates the agent runtime, inference engine, and third-party components into independently attested confidential virtual machines and mediates their interaction through secure communication channels, protecting the confidentiality and integrity of all stakeholders. Our evaluation demonstrates that this approach is practical, achieving strong isolation guarantees with less than 5.15\% runtime overhead compared to standard OS process-level deployments.

\section{Acknowledgment}
We thank the anonymous reviewers for their insightful comments and suggestions.  The research was supported by the UKRI  Open Plus Fellowship (EP/W005271/1, Securing the Next Billion Consumer Devices on the Edge), the Amazon Research Award “Auditable Model Privacy using TEEs”, and the AI Security Institute (AISI) Systemic Safety Grants Programme (UKRI833). 

Professor Kotz was supported by a Royal Society Wolfson Visiting Fellowship and by a collaborative award from the U.S. National Science Foundation (NSF) SaTC Frontiers program under award number 1955805. 

The views and conclusions contained herein are those of the authors and should not be interpreted as necessarily representing the official policies, either expressed or implied, of any sponsor. Any mention of specific companies or products does not imply any endorsement by the authors, by their employers, or by their sponsors.

\bibliographystyle{ACM-Reference-Format}
\bibliography{my_bib}

\end{document}